\documentclass[12pt]{article}
\usepackage{cite}
\usepackage[pdftex]{graphicx}
\usepackage[cmex10]{amsmath}
\usepackage{bm}
\usepackage{epstopdf}
\usepackage{amsthm}
\usepackage{parskip}

\usepackage{amssymb} % Added by Jesse!!

\theoremstyle{definition}

\usepackage{caption}
\usepackage{subcaption}
\usepackage{url}
\newcommand{\defn}[1]{\emph{#1}}
\newcommand{\boundary}{\partial}
\newcommand{\set}[1]{\left\{#1\right\}}

\begin{document}
\interdisplaylinepenalty=2500

\title{An application of topological graph clustering to protein
  function prediction}

% \author{
%     \IEEEauthorblockN{Author1\IEEEauthorrefmark{1}, Author2\IEEEauthorrefmark{2}, Author3\IEEEauthorrefmark{2}, Author4\IEEEauthorrefmark{1}}
%     \IEEEauthorblockA{\IEEEauthorrefmark{1}Institution1
%     \\\{1, 4\}@abc.com}
%     \IEEEauthorblockA{\IEEEauthorrefmark{2}Institution2
%     \\\{2, 3\}@def.com}
% }

\author{R.\ Sean Bowman\\Jesse Johnson \and Douglas Heisterkamp\\Danielle O'Donnol}

% \author{
%   \IEEEauthorblockN{R. Sean Bowman\IEEEauthorrefmark{1}, Douglas
%     Heisterkamp\IEEEauthorrefmark{2}, Jesse
%     Johnson\IEEEauthorrefmark{1}, Danielle
%     O'Donnol\IEEEauthorrefmark{1}}
%   \IEEEauthorblockA{\IEEEauthorrefmark{1}
% Department of Mathematics\\
% Oklahoma State University\\
% sean.bomwan@okstate.edu\\
% jjohnson@math.okstate.edu\\
% odonnol@math.okstate.edu}
%   \IEEEauthorblockA{\IEEEauthorrefmark{2}
% Department of Computer Science\\
% Oklahoma State University\\
% doug@cs.okstate.edu}}

% \author{
% \IEEEauthorblockN{Sean Bowman}
% \IEEEauthorblockA{Department of Mathematics\\
% Oklahoma State University\\
% sean.bomwan@okstate.edu}
% \and
% \IEEEauthorblockN{Douglas Heisterkamp}
% \IEEEauthorblockA{Department of Computer Science\\
% Oklahoma State University\\
% doug@cs.okstate.edu}
% \and
% \IEEEauthorblockN{Jesse Johnson}
% \IEEEauthorblockA{Department of Mathematics\\
% Oklahoma State University\\
% jjohnson@math.okstate.edu}
% \and
% \IEEEauthorblockN{Danielle O'Donnol}
% \IEEEauthorblockA{Department of Mathematics\\
% Oklahoma State University\\
% odonnol@math.okstate.edu}
% }

\maketitle

\begin{abstract}
  We use a semisupervised learning algorithm based on a topological
  data analysis approach to assign functional categories to yeast
  proteins using similarity graphs.  This new approach to analyzing
  biological networks yields results that are as good as or better
  than state of the art existing approaches.
\end{abstract}

% \begin{IEEEkeywords}
% Clustering, Graph Partitioning, Pinch Ratio 
% \end{IEEEkeywords}

\section{Introduction}
Determining protein function is an integral part of understanding
biological mechanisms.  Until recently, proteins were characterized
one at a time, slowly building our knowledge of a given pathway or
mechanism.  The yeast \emph{Saccharomyces cerevisiae} was one of the
first organisms to have its genome sequenced~\cite{Goffeau96}.  As a
model organism it has been thoroughly studied.  However, at the time of
sequencing only about a third of the genes had been characterized.
With the increasing ease of DNA sequencing, databases were made of
proteins with known function from different organisms.  Sequences
could be compared with those in the database, and those with sequence
similarity would likely correspond to a similar protein in a different
organism.  For \emph{S. cerevisiae}, about one third showed sequence
similarity to genes in other organisms, leaving the last third,
approximately 2000 genes, completely unknown.  The number of unknown
genes and difficulty in understanding their function called for a new
approach.

The shift to large scale biology has brought about both necessity for
large scale protein function identification and many new avenues for
large scale protein characterization.  There are now hundreds of
genomes sequenced.  Before, being able to sequence a gene was a
major obstacle.  Now, the difficulty has shifted to determining the
function of a large number of genes.  There are numerous new high
throughput methods producing information about all different
attributes of a protein, like the Pfam domain, protein-protein
interaction, protein expression, etc.  This gives us a more complete
picture of these complex systems, and also gives us vast amounts of
information for which there are not always obvious conclusions.  For
more background information, see~\cite{Twyman04} and~\cite{Kraj08}.

We can represent information about different proteins in the form of a
graph.  For example, with protein-protein interaction, the proteins
are represented by vertices and edges between vertices represent
interactions.  Given such a graph, it is natural to apply approaches
from machine learning to find clusters of vertices.  The idea is that
proteins that form a tightly grouped cluster in such a graph will have
similar functions.  Thus we can predict the function of an unknown
protein when there are proteins of known function in its cluster.
This strategy has been successfully implemented by many groups.  In
this paper, we apply a novel semisupervised topological data analysis
approach to the same problem.  This algorithm is based on the TILO/PRC
(Topologically Intrinsic Lexicographic Ordering/Pinch Cluster Ratio)
algorithm developed by the second and third
author~\cite{Johnson12,HeisterkampJohnson12} and described in the next
section.

Many approaches to functional annotation given a protein interaction
graph have been proposed~\cite{Sharan07}.  Some have focused on
techniques from statistics and machine learning such as Markov random
field models~\cite{DengEtAl} and Support Vector Machines
(SVM)~\cite{LanckrietEtAl}.  Clustering using TILO/PRC bears some
similarity to Spectral clustering~\cite{ShiMalik00} in that it
attempts to find a partition of a graph with small boundary but large
interior weight.  Our approach is therefore similar in spirit to
methods that make use of the graph Laplacian in clustering or
propagation of
labels~\cite{TrivodalievEtAl11,ShinEtAl,SenEtAl06,Tran}.  Note that
several of these methods use a combined graph obtained as a weighted
average of several other graphs.  The weights are obtained as part of
the procedure using, for example, semidefinite programming
in~\cite{LanckrietEtAl}.  We have not attempted to do this in the
present work, and use a simple average to form a combined graph.  As
we will see, our results are on par with methods that attempt to
select optimal weights with regard to some objective function.

%% reminiscent of Spectral Clustering, but does
%% not rely on the theoretical relation between the Cheeger constant and
%% eigenvectors of the graph
%% Laplacian~\cite{Johnson12,HeisterkampJohnson12}. In addition to a
%% partition of the data set, TILO/PRC also returns a set of values
%% called \textit{pinch ratios} (the PR in PRC) that we can think of as
%% measuring how well separated each cluster is from the rest of the data
%% set.

%% In particular, the pinch ratio for a cluster is zero if it is totally
%% isolated, i.e.\ there is no edge from a vertex in this cluster to a
%% vertex outside the cluster. A pinch ratio of 1 means that the cluster
%% is geometrically indistinguishable from the rest of the data
%% set. (Unless told otherwise, TILO/PRC will only return clusters with
%% pinch ratios strictly less than 1.)

The contributions of this paper are as follows:

\begin{enumerate}
\item We describe the TILO algorithm for semisupervised learning.
\item We demonstrate that this algorithm is effective at predicting
  protein function.
\item We demonstrate that this algorithm performs well in comparison
  to existing semisupervised methods, including the Markov random
  field model of Deng, Chen, and Sun~\cite{DengEtAl}, and the kernel
  methods of Lanckriet et al.~\cite{LanckrietEtAl}.
\end{enumerate}

Code for the project is located at
\url{http://seanbowman.me/yeast-protein}.

\section{The TILO algorithm}
First we describe the TILO/PRC algorithm, which clusters the vertices
of a graph by finding a linear order on the vertices with nice
properties.  See~\cite{Johnson12,HeisterkampJohnson12} for more
details.  Let $G=(V, E)$ be a graph, where $V$ is the set of vertices
and $E$ is a set of weighted edges.  Given a set of vertices
$A\subseteq V$, define the \defn{boundary} $\boundary A\subseteq E$ to
be the set of edges with one endpoint in $A$ and the other in
$V\setminus A$.  The \defn{size} of the boundary, $|\boundary A|$, is
defined to be the sum of weights of $\boundary A$.

A \defn{pinch cluster} in $G$ is a set of vertices $S\subseteq V$ such
that for any sequence of vertices $v_1,\dots,v_m$, if adding these
vertices to $S$ or removing them from $S$ creates a set with smaller
boundary, then for some $k<m$, adding/removing $v_1,\dots, v_k$
to/from creates a set with strictly larger boundary.  Informally, a
pinch cluster has small boundary relative to the edges it contains,
and vertices in a pinch cluster are more strongly connected to each
other than to outside vertices.  

Consider an ordering $O=(v_1, v_2,\dots, v_n)$ of $V$.  Let
$A_i=\set{v_1,\dots, v_i}$ and $b_i = |\boundary A_i|$.  The
\defn{width} of $O$, $w(O)$ is the tuple $(w_1,\dots,w_{n-1})$, where
$\set{w_1,\dots, w_{n-1}} = \set{b_1,\dots, b_{n-1}}$ and $w_{i+1}\leq
w_i$ for each $1\leq i< n-1$.  We order widths lexicographically.
In~\cite{Johnson12} a property called \defn{weak reducibility} is defined
for an ordering $O$, and~\cite[Lemma 3]{Johnson12} shows that given a
weakly reducible ordering $O$, we can find a new ordering $O'$ so that
$w(O')<w(O)$.  After applying this procedure a finite number of times,
we arrive at a \defn{strongly irreducible} ordering.

The goal of the TILO algorithm is to find a strongly irreducible
ordering of the vertices of $G$.  Suppose that $O$ is a strongly
irreducible ordering and $i$ is the index of a local minimum of
$(b_1,\dots, b_{n-1})$.  Then~\cite[Theorem 4]{Johnson12} shows that
$A_i$ and its complement are pinch clusters.  Note that we are free to
pick the index of \emph{any} local minimum of the boundary vector in order to
partition the data.  The pinch ratio cut, defined
in~\cite{HeisterkampJohnson12}, is one heuristic for choosing such an
index which has been shown to work well in practice.  However, in the
present work we will always make cuts at \emph{every} index
corresponding to a local minimum.

%% The pinch
%% ratio cut is a criterion for choosing such an
%% index~\cite{HeisterkampJohnson12}.  The pinch ratio at index $k$ is
%% defined to be
%% \[ \frac{b_k}{\min\left(\max_{v_i\in A_k} b_i, \max_{v_j\in V\setminus
%%       A_k} b_j\right)}. \]

%% This is a measure of XXX In~\cite{HeisterkampJohnson12} it was shown
%% that a clustering algorithm based on pinch ratios offers performance
%% comparable to or better than established algorithms such as K Means
%% and Spectral clustering.

\subsection{Semisupervised learning with TILO}
In the case at hand, we are given a weighted graph $(V,E)$ which we
think of as a similarity graph.  The vertex set
$V=\set{v_1,v_2,\dots,v_n}$ is divided into two parts: $m$ vertices
$v_1,\dots,v_m$ are labeled $y_1,\dots,y_m$, and the rest are
unlabeled.  We wish to use the structure of the graph to assign labels
to $v_{m+1},\dots,v_n$.  In the present application $y_i\in\set{0,1}$
for $1\leq i\leq m$, and so we may also assign probabilities $y_i\in
[0,1]$, $m<i\leq n$, representing the likelihood that vertex $v_i$ has
the label $1$.

For each connected component of the similarity graph, we use the
TILO/PRC algorithm to divide the vertices into clusters.  Given a
strongly reducible ordering $O$, we make a cut at every local minimum
of the boundary vector, thus cutting the graph into the maximal number
of pinch clusters.  If $v_{i_1},v_{i_2},\dots,v_{i_k}$ are labeled
vertices in a resulting cluster, we assign the probability
\[ \frac{1}{k}\sum_{j=1}^{k} y_{i_j} \]
to each unlabeled vertex in the cluster.  If there are no labeled
vertices in the cluster, we assign the most common label in the data
set to each unlabeled vertex.

The TILO algorithm can get stuck in local minima, which correspond to
orderings that may realize suboptimal clusters in a given graph.  We
use a version of bagging~\cite{Breiman96} in order to generate more
accurate predictions from our labeled data.  Bagging also ensures that
the probability that a given vertex always receives the most common
label in the data set is extremely low.  After generating $N$ subsets
of the unlabeled data by sampling a proportion $\lambda$ uniformly
without replacement, the algorithm described above is run on each
sample.  The probabilities obtained are averaged over all runs, and
the averaged probabilities $y_{m+1},\dots,y_n$ are the output of the
algorithm.  In the data below, we use $N=25$ and $\lambda=0.5$. 

\section{Experimental Results}
We use the data set from~\cite{Tran} and also used in~\cite{DengEtAl}
and~\cite{LanckrietEtAl} (among others) in order to compare our
results.  The goal of the algorithm is to predict functional classes
of 3588 yeast proteins, with true labels given by the MIPS
Comprehensive Yeast Genome Database.  There are 13 classes in the top
level hierarchy, and we view the problem as 13 separate binary
classification tasks.

The input data consists of five symmetric matrices $W_i$,
$i=1,2,3,4,5$, whose entries describe different types of interaction
between the row and column proteins.
The first matrix, $W_1,$ comes from the Pfam domain of the proteins.  
Each protein was characterized based on the presence or absence of 3950 different structural domains.  
This results in 3950-bit vector, and the dot products of these vectors
are used to generate the matrix.  
The next three, $W_2, W_3,$ and $W_4,$ are from the combined data in
the MIPS Comprehensive Yeast Genome Database.  
The matrix $W_2$ indicates if there is co-participation in a protein complex.  
The matrix $W_3$ indicates known protein-protein interactions.  
The matrix $W_4$ indicates known genetic interactions.  
The final matrix $W_5$ is based on comparing the proteins' gene expression profiles.  
%Considered "the same" i.e. entry of 1 if expression profiles exhibits a Pearson correlation greater than 0.8. 

The graphs obtained by considering the $W_i$ as adjacency matrices are
highly disconnected and have many components with a small number of
vertices.  We do not consider isolated vertices, since it is
impossible to infer a label using our graph based approach.
Furthermore, $W_5$ is so sparse that we have not used it in our tests.
We also examine the integrated graph
\[ \frac{1}{5}\sum_{i=1}^5 W_i. \]
This graph is relatively well connected.  We use 5 fold cross
validation 3 times and report the receiver operating characteristic
(ROC) score on the test set.

Our results are shown in Table~\ref{table1}.  Shown are the mean ROC
scores, as described above, for the graphs $W_1$, $W_2$, $W_3$, $W_4$,
and the integrated graph.  The graph $W_1$ has 432 components, 2809
vertices, and 48445 edges; $W_2$ has 29 components, 1051 vertices, and
1872 edges; $W_3$ has 140 components, 1342 vertices, and 1844 edges;
$W_4$ has 106 components, 819 vertices, and 1006 edges; the integrated
graph has 96 components, 3278 vertices, and 56371 edges.  The columns
labeled SVM/$W_1$ and SDP/SVM are results from~\cite{LanckrietEtAl}.
The first shows the ROC values obtained using a $1$--norm SVM with
$C=1$ on a normalization of our matrix $W_1$, and the SDP/SVM column
shows the performance of a semidefinite programming SVM approach which
uses a weighted combination of the $W_i$.  The last column shows the
performance of the Markov random field algorithm of~\cite{DengEtAl} on
the weighted combination used in the previous column.

%% Following the suggestion of~\cite{SantosEmbrechts}, we also report the
%% Adjusted Rand Index (ARI) of an associated clustering.  Suppose that
%% we are given a graph with $m$ labeled vertices and $n-m$ unlabeled
%% ones.  Recall that the vertices are labeled $y_1,\dots, y_m$, and the
%% algorithm outputs values $y_{m+1},\dots,y_n$ in the interval
%% $[0,1]$.  We determine the number $c$ minimizing
%% \[ \left| \frac{\#\allsuchthat{y_i}{y_i<c, i=m+1,\dots, n}}{n-m}
%%   - \frac{\#\allsuchthat{y_i}{y_i=0, i=1,\dots,m}}{m} \right| \]
%% and then set new labels for the test set according to
%% \[ \hat{y}_i = \begin{cases}
%%   0 & \text{if $y_i<c$,}\\
%%   1 & \text{otherwise.}
%% \end{cases} \]
%% In other words, we find a value of $c$ for which the proportion of
%% unlabeled points with $y_i$ less than $c$ is approximately equal to
%% the proportion of labeled points with label zero.  We then compute the
%% ARI of the $\hat{y}_i$ versus the true labels.  

\begin{table}[h]
  \centering
  \[ \begin{array}{ccccc}
    \text{Class} & \text{W1} & \text{W2} & \text{W3} &
    \text{W4}\\\hline
    1 & 0.880 \pm 0.019 & 0.768 \pm 0.046 & 0.700 \pm 0.027 & 0.833 \pm 0.043\\
    2 & 0.879 \pm 0.031 & 0.684 \pm 0.098 & 0.722 \pm 0.059 & 0.756 \pm 0.165\\
    3 & 0.825 \pm 0.022 & 0.793 \pm 0.040 & 0.755 \pm 0.033 & 0.881 \pm 0.029\\
    4 & 0.876 \pm 0.014 & 0.844 \pm 0.020 & 0.842 \pm 0.021 & 0.857 \pm 0.034\\
    5 & 0.896 \pm 0.019 & 0.847 \pm 0.032 & 0.816 \pm 0.058 & 0.830 \pm 0.091\\
    6 & 0.881 \pm 0.021 & 0.751 \pm 0.046 & 0.767 \pm 0.039 & 0.830 \pm 0.023\\
    7 & 0.878 \pm 0.022 & 0.863 \pm 0.042 & 0.887 \pm 0.027 & 0.866 \pm 0.039\\
    8 & 0.794 \pm 0.034 & 0.656 \pm 0.103 & 0.651 \pm 0.052 & 0.776 \pm 0.067\\
    9 & 0.823 \pm 0.054 & 0.589 \pm 0.149 & 0.811 \pm 0.063 & 0.744 \pm 0.079\\
    10& 0.767 \pm 0.034 & 0.755 \pm 0.045 & 0.776 \pm 0.045 & 0.804 \pm 0.025\\
    11& 0.629 \pm 0.050 & 0.625 \pm 0.056 & 0.674 \pm 0.055 & 0.635 \pm 0.070\\
    12& 0.961 \pm 0.019 & 0.756 \pm 0.109 & 0.796 \pm 0.056 & 0.759 \pm 0.126\\
    13& 0.854 \pm 0.053 & 0.577 \pm 0.159 & 0.710 \pm 0.104 & 0.753 \pm 0.107\\
  \end{array} \]
  \caption{Mean ROC of the semisupervised TILO/PRC algorithm on the
    graphs $W_i$.}\label{table1}
\end{table}

\begin{table}[h]
  \centering
  \[ \begin{array}{ccccc}
   
    \text{Class} & \text{Integrated} & \text{SVM}/W_1 & \text{SDP/SVM}
    & \text{MRF}\\\hline
    1 & 0.866 \pm 0.018 & .8373\pm.0037 & .8825\pm.0042 & .7532\pm.0042\\
    2 & 0.871 \pm 0.033 & .8107\pm.0113 & .8563\pm.0100 & .7173\pm.0102\\
    3 & 0.839 \pm 0.019 & .7547\pm.0062 & .8464\pm.0053 & .6990\pm.0045\\
    4 & 0.882 \pm 0.011 & .8085\pm.0048 & .9024\pm.0028 & .7409\pm.0049\\
    5 & 0.908 \pm 0.014 & .8349\pm.0058 & .9094\pm.0050 & .7375\pm.0102\\
    6 & 0.868 \pm 0.022 & .8069\pm.0046 & .8742\pm.0049 & .7183\pm.0059\\
    7 & 0.880 \pm 0.019 & .8010\pm.0074 & .9149\pm.0040 & .7534\pm.0085\\
    8 & 0.811 \pm 0.030 & .7023\pm.0089 & .8023\pm.0094 & .7285\pm.0120\\
    9 & 0.823 \pm 0.043 & .7309\pm.0113 & .8623\pm.0091 & .6849\pm.0107\\
    10& 0.790 \pm 0.023 & .6906\pm.0079 & .8120\pm.0078 & .6954\pm.0060\\
    11& 0.658 \pm 0.043 & .5952\pm.0148 & .6575\pm.0093 & .5691\pm.0092\\
    12& 0.954 \pm 0.018 & .9331\pm.0038 & .9674\pm.0023 & .8575\pm.0076\\
    13& 0.841 \pm 0.068 & .6678\pm.0227 & .8083\pm.0091 & .6612\pm.0169
  \end{array} \]

  \caption{Mean ROC of the semisupervised TILO/PRC algorithm on the
    integrated graph.  Columns SVN/$W_1$ and SDP/SVM are results
    of~\cite{LanckrietEtAl}, and the last column is results
    from~\cite{DengEtAl}.  See text for details.}\label{table2}
\end{table}

\section{Conclusion}

Understanding protein function is integral to our understanding of
pathways and mechanisms in biological systems, and the huge amount of
data available today necessitates techniques which can work on a large
scale.
We have described a semisupervised learning algorithm based on
TILO/PRC and shown that this algorithm performs well on the task of
predicting protein functional classes on a data set of yeast proteins.

Tables~\ref{table1} and~\ref{table2} show that for each of the 13
functional classes studied, the semisupervised TILO method performs
better than the SVM method of~\cite{LanckrietEtAl} when using $W_1$.
The same is true for $W_2$, $W_3$, and $W_4$.  In all, the mean ROC
score improves from 0.767 to 0.842 on $W_1$, albeit with slightly
higher standard deviations.  For the combined graph, we obtain similar
scores (a mean ROC of 0.844 versus 0.853 for the SDP/SVM approach
of~\cite{LanckrietEtAl}) while using a simple average of the $W_i$
instead of a weighted average.

We believe that algorithms inspired from ideas in topology, such as
the TILO/PRC algorithm described here, are uniquely suited to this
problem for several reasons.  They have a good theoretical grounding
with clear links to both topology and existing machine learning
approaches, they operate directly on protein interaction graphs, they
are less sensitive to geometric properties of the data (which are not
always relevant), and they are capable of meeting the challenges of
large scale biology described above.

\textbf{Acknowledgments:} we would like to thank Loc Tran for
providing us with the data used in our experiments.

\clearpage
\bibliographystyle{plain}
\bibliography{yeast-protein}

\end{document}